\documentclass[preprint,prl,onecolumn]{revtex4}
\usepackage{latexsym}
\usepackage{amssymb}
\usepackage{amsmath}
\usepackage{graphicx}

\begin{document}

\title{Observation of an in-plane magnetic-field-driven phase transition in
a quantum Hall system with SU(4) symmetry }
\author{G. P. Guo$^{(1)}$, Y. J. Zhao$^{(1)}$, T. Tu$^{(1)\text{ }\ast }$,
X. J. Hao$^{(1)}$, X. C. Zhang$^{(2)}$, G. C. Guo$^{(1)}$ and H. W. Jiang$%
^{(2)}$ }
\email{jiangh@physics.ucla.edu,tuao@ustc.edu.cn}
\affiliation{$^{(1)}$ Key Laboratory of Quantum Information, University of Science and
Technology of China, Chinese Academy of Sciences, Hefei 230026, P. R. China\\
$^{(2)}$ Department of Physics and Astronomy, University of California at
Los Angeles, 405 Hilgard Avenue, Los Angeles, CA 90095, USA }
\maketitle

%%%%%%%%%%%%%%%%%%%%%%%%%%%%main body%%%%%%%%%%%%%%%%%%%%%%%%%%%%
\baselineskip20pt

\textbf{In condensed matter physics, the study of electronic states with
SU(N) symmetry has attracted considerable and growing attention in recent
years, as systems with such a symmetry can often have a spontaneous
symmetry-breaking effect giving rise to a novel ground state. For example,
pseudospin quantum Hall ferromagnet of broken SU(2) symmetry has been
realized by bringing two Landau levels close to degeneracy in a bilayer
quantum Hall system \cite{DasSarma}. In the past several years, the
exploration of collective states in other multi-component quantum Hall
systems has emerged \cite{Hirayama,Jiang2005,Tsui,Shayegan}. Here we show
the conventional pseudospin quantum Hall ferromagnetic states with broken
SU(2) symmetry collapsed rapidly into an unexpected state with broken SU(4)
symmetry, by in-plane magnetic field in a two-subband GaAs/AlGaAs
two-dimensional electron system at filling factor around $\nu=4$. Within a
narrow tilting range angle of $0.5$ degrees, the activation energy increases
as much as $12$ K. While the origin of this puzzling observation remains to
be exploited, we discuss the possibility of a long-sought pairing state of
electrons with a four-fold degeneracy.}

The studies of multi-component quantum Hall systems so far have been limited
to systems with a SU(2) symmetry. Very recently, interest has been further
extended to systems with a higher order symmetry, such as SU(4), motivated
mainly by the surge of research in graphene, where the two-fold spin and
two-fold valley degeneracy lead to a four fold degeneracy \cite%
{graphene1,graphene2}. The SU(4) symmetry can be readily created in a
multi-component quantum Hall system. For a typical two subband semiconductor
heterostructure, where both the symmetric and anti-symmetric subbands of the
confined quantum well are occupied, two distinct sets of Landau levels are
present. Through varying the density or magnetic field, levels with
different Landau orbital indices originating from the two subbands can be
brought into degeneracy. Due to the very small energy difference of the
spin-splitting in GaAs \cite%
{Wescheider,Hirayama,Jiang1999,Jiang2005,Jiang2006}, the system can provide
us with the desired SU(4) symmetry. The experimental studies presented in
this paper are specifically focused on this interesting physical regime.

The sample was grown by molecular-beam epitaxy and consists of a symmetrical
modulation-doped $24$ nm wide single GaAs quantum well bounded on each side
by Si $\delta $-doped layers of AlGaAs with doping level $n_{d}=10^{12}$ cm$%
^{-2}$. Heavy doping creates a very dense 2DEG, resulting in the filling of
two subbands in the well. As determined from the Hall resistance data and
Shubnikov-de Haas oscillations in the longitudinal resistance, the total
density is $n=8.0\times 10^{11}$ cm$^{-2}$, where the first and the second
subband have a density of $n_{1}=6.1\times 10^{11}$ cm$^{-2}$ and $%
n_{2}=1.9\times 10^{11}$ cm$^{-2}$. The sample has a low-temperature
mobility $\mu =4.1\times 10^{5}$ cm$^{2}$/V s, which is extremely high for a
2DEG with two filled subbands. The samples are patterned into Hall bars
using standard lithography techniques. A NiCr top gate is evaporated on the
top of the sample, approximately $350$ nm away from the center of the
quantum well. By applying a negative gate voltage on the NiCr top gate, the
electron density can be varied continuously. Magneto-transport measurements
were carried out in an Oxford Top-Loading Dilution Refrigerator with a base
temperature of $15$ mK and a \textit{in situ} motorized rotating stage with
a resolution better than $0.01$ degree. To measure the longitudinal and Hall
resistance $R_{xx}$ and $R_{xy}$, we used a standard ac lock-in technique
with electric current ranging from $10$ nA to $100$ nA at a frequency of $%
11.3$ Hz. Two devices from the same wafer were studied, and they have
produced remarkably identical results. For consistency, however, we present
the data from only one sample.

In this experiment, we have concentrated our study around $\nu=4$, where
four energy levels are filled. In the absence of an in-plane magnetic field
(i.e., with zero tilting angle), we have essentially reproduced the results
of earlier studies \cite{Jiang2005,Jiang2006}on devices from the same wafer.
The topology of longitudinal resistance $R_{xx}$ in the density($n$%
)-magnetic field($B$) plane exhibits a square-like structure around $\nu=4$,
as shown in Fig. 1a. Here, point B corresponds to the degeneracy point of $%
\left\vert (S,1,\downarrow )\right\rangle $ and $\left\vert (A,0,\uparrow
)\right\rangle $ while point C corresponds to that of $\left\vert
(S,1,\uparrow )\right\rangle $ and $\left\vert (A,0,\downarrow
)\right\rangle $, as illustrated schematically in Fig.(1f). Here we label
the single-particle levels ($i,N,\sigma $), and $i$($=S,A$), $N$, and $%
\sigma $($=\uparrow ,\downarrow $) are the subband, orbital and spin quantum
numbers. One prominent feature is the disappearance of the extended states
(i.e., bright lines) that complete the two-arms of the square. The
disappearance is due to the pseudospin gaps of easy-axis quantum Hall
ferromagnetic states where the electrons are suddenly transferred from ($%
S,1,\downarrow $) to ($A,0,\uparrow $), and from ($S,1,\uparrow $) to ($%
A,0,\downarrow $), or vice versa \cite{Hirayama,Jiang2006,Macdonald2000}.

\begin{figure}[h]
\includegraphics[width=0.8\columnwidth]{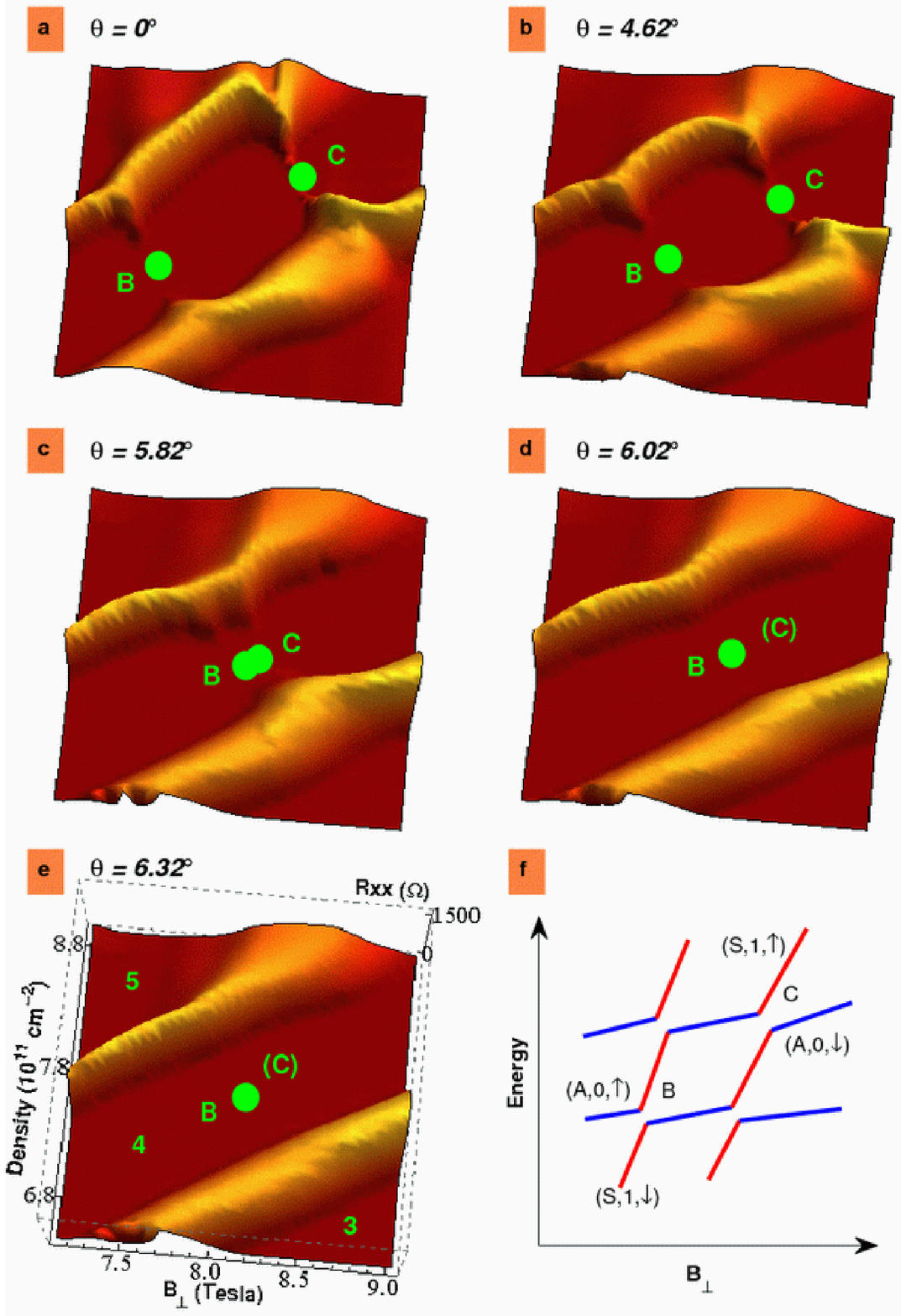}
\caption{Topology structures of the longitudinal resistance $R_{xx}$ in $%
n-B_{\perp }$ phase diagram at filling factor $\protect\nu =4$ with the
tilted angle $\protect\theta $ from $0^{\circ }$ to $7^{\circ }$, which are
measured at the base temperature. (a)-(e): phase diagram at tilted angle $%
\protect\theta =0^{\circ },\ 4.62^{\circ },5.82^{\circ },\ 6.02^{\circ },\
6.32^{\circ }$. (f): schematic drawing of the anti-crossing between
different subband and spin indices Landau levels in two corresponding places
marked by B and C in fig.(a)-(e). }
\label{phasediagramtilted}
\end{figure}

Now we turn our attention to the behavior of the topology phase diagram in
the tilted magnetic field. The density($n$)-perpendicular magnetic field($%
B_{\bot }$) phase diagrams of $R_{xx}$ at several tilted angles $\theta
=0^{\circ }$,\thinspace $4.62^{\circ }$, $5.82^{\circ }$, $6.02^{\circ }$, $%
6.32^{\circ }$ for $\nu=4$ are shown in Fig.(\ref{phasediagramtilted}). Here
the angle $\theta $ is defined as $\tan \theta =\frac{B_{\Vert }}{B_{\bot }}$%
. When the titling angle is increased to $4.62^{\circ }$, Fig.(1b), the size
of the square, a measure of the energy separation between the two degeneracy
points, shrinks only slightly. However, to increase merely another 1-degree,
the two "arms" of the square structure almost collapse together, Fig.(1c),
and become one point at slightly larger angles, Fig.(1d),(1e). The evolution
of the positions of the anti-crossing points B and C (roughly in the middle
of each of the "arms" of the square structure at $\nu=4$, also the local
minimum of energy gap in the anti-crossing region), are plotted in Fig.(\ref%
{degeneracy}) as a function of the tilting angle (See also the movie in the
supplement). One can clearly visualize that the two points get close with
each other in a narrow range of tilting angle and emerge as one point when $%
\theta $ is increased, which implies four Landau levels are brought into
degeneracy. In other words, the two degeneracy points with a SU(2) symmetry
become a single degeneracy point with a SU(4) symmetry.

Since there seems to be a drastic change in the characteristics of the phase
diagram, in a very narrow range of the in-plane magnetic field, it is
natural to obtain a measurement of the energy scale of the associated energy
gaps there. In Fig.(\ref{energygap}), we present the $\theta $ dependence of
the energy gap $E_{a}$, deduced from the thermal activation relation $\rho
_{xx}\sim \exp (-E_{a}/2T)$, at two points B, C marked by the dots in Fig.(%
\ref{phasediagramtilted}). The energies are found to be nearly constant, $%
\approx 3$ K before $\theta =5.82^{\circ }$, while increasing sharply for a
small angle region into a rough angle independent region $\approx 15$ K.
Without any assist of theoretical analysis, the raw data strongly suggests a
phase transition.

\begin{figure}[h]
\includegraphics[width=0.8\columnwidth]{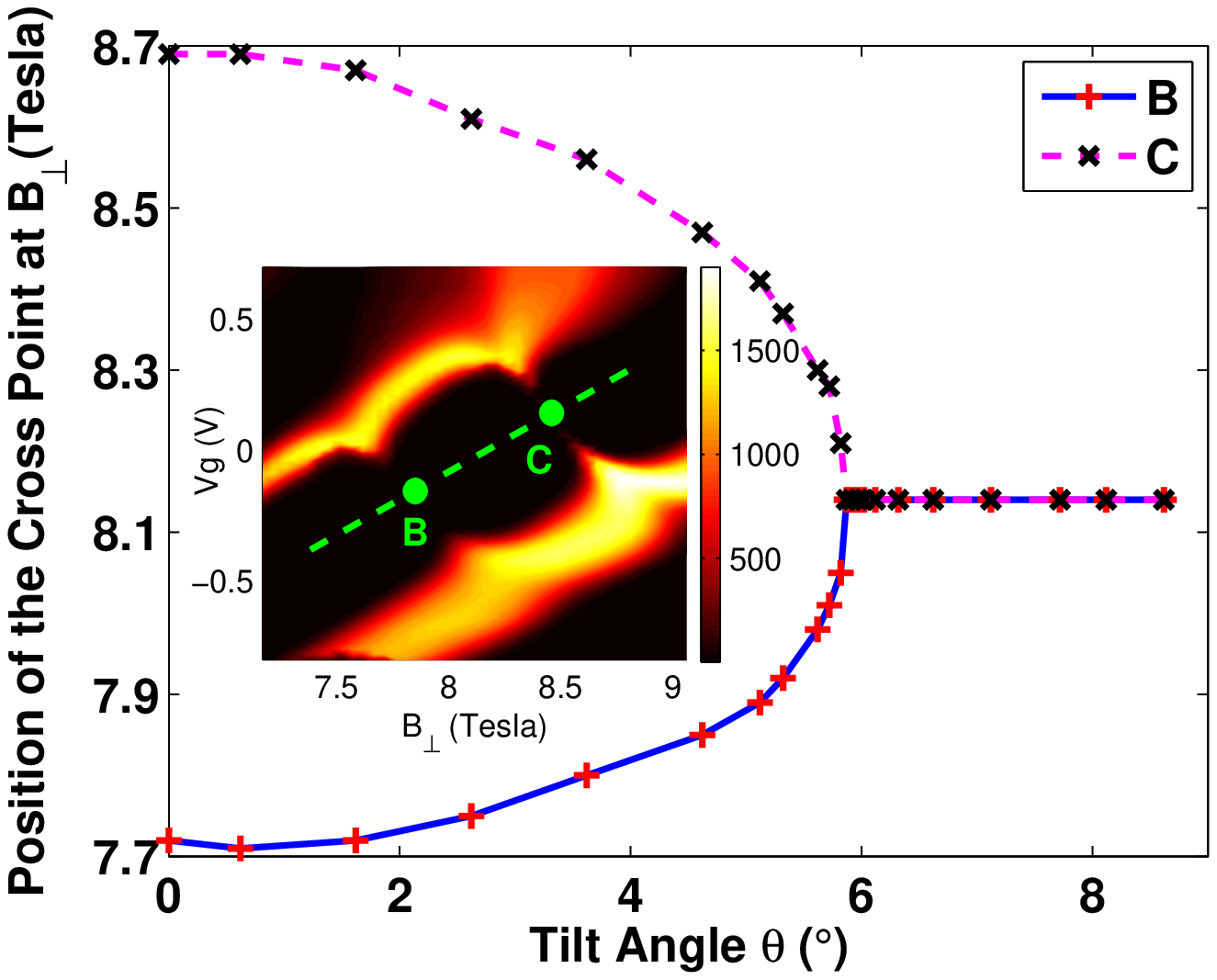}
\caption{Evolution of the positions of the degenerate points B and C as a
function of the tilted angle $\protect\theta $. The corresponding in-plane
magnetic field are displayed on the top axis. Insert: indication of the
points that are measured.}
\label{degeneracy}
\end{figure}

Before we speculate on the physical origin of the observations, we would
like to compare the current experiment with others in similar 2DEG systems
in the presence of an in-plane magnetic field. There are indeed examples of
in-plane magnetic field induced phase transitions in multi-component quantum
Hall systems. The system studied by Murphy \textit{et al. }\cite{Eisenstein}%
, a strongly coupled double quantum well at a filling factor $\nu=1$, showed
a relatively large energy gap change when the tilted angle was increased. It
is now commonly believed that this change reflects a commensurate to an
incommensurate phase transition in the pseudospin field. However, in this
case, only the two lowest Landau levels are involved before and after this
quantum phase transition, in contrast to four Landau levels degeneracy in
our system. In a recent study of Si-SiGe heterostructures by Lai \textit{et
al.} \cite{Tsui}, the $\nu=3$ and $5$ valley gaps rise rapidly when the
angle of tilting approaches about $65^{\circ }$. In this experiment, the
effect is driven by the coincidence of two different Landau levels when the
Zeeman splitting becomes comparable to the cyclotron energy, which is not
possibly consistent with our small tilting angle. In another recent study of
the 2DEG in a AlAs quantum well by Vakili \textit{et al.} \cite{Shayegan},
the size of the energy gap of $\nu=3$ valley changes rapidly through a
coincidence angle of about $65^{\circ }$. The change is induced also by the
crossing of two different Landau levels, made possible again by the large
tilting angle.

\begin{figure}[h]
\includegraphics[width=0.9\columnwidth]{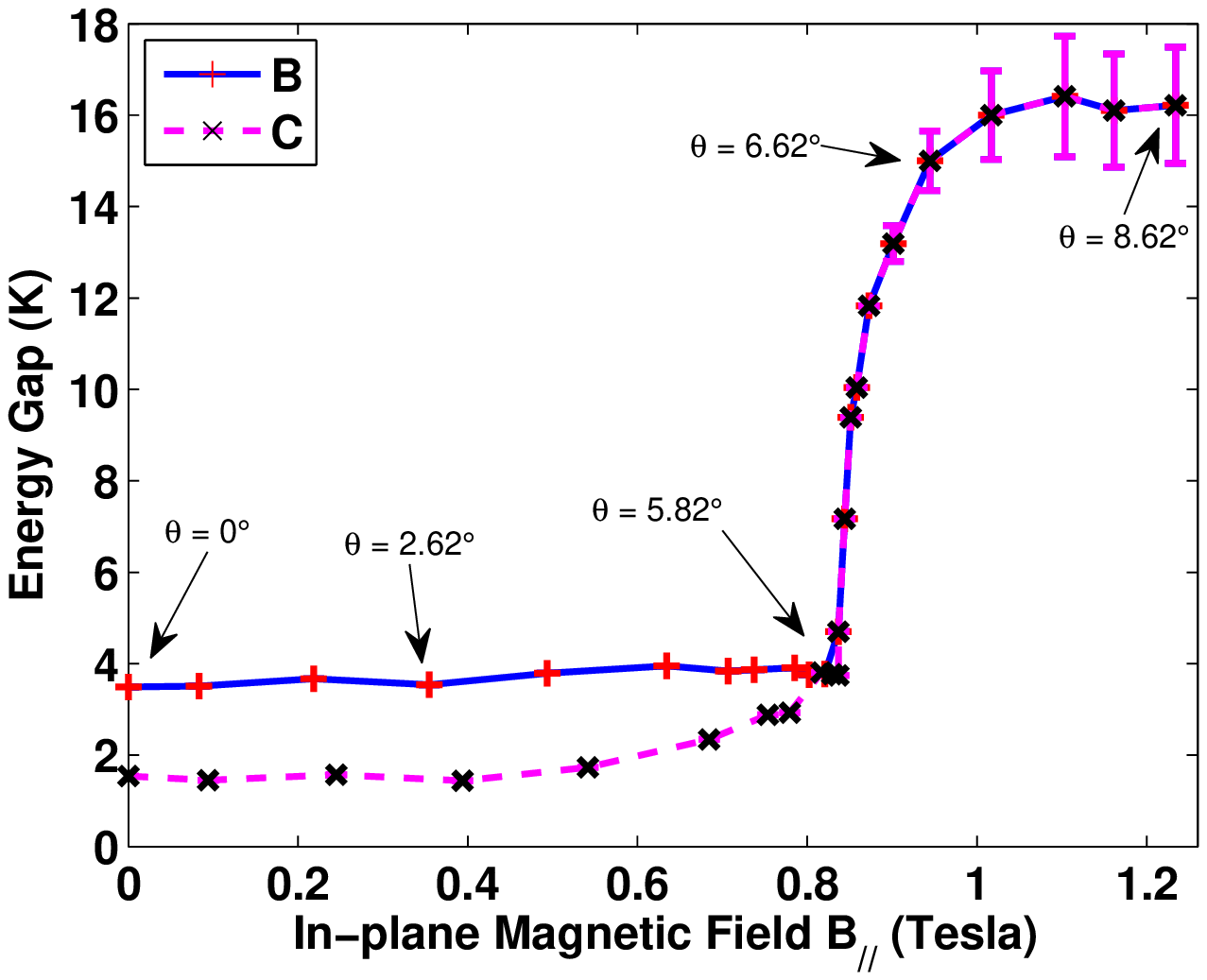}
\caption{Energy gap of two anti-crossing points B and C as a function of
in-plane magnetic field $B_{\parallel}$.}
\label{energygap}
\end{figure}

Changes in the energy gap $E_{a}$ normally reflect changes in the spectrum
of charged excitations and the 2DEG ground state. At first sight, one might
expect that the phase transition is induced by the spin and pseudospin flips
due to variations in the Zeeman energy $g\mu _{B}B_{tot}$. The Zeeman
energy, however, changes only about 30 mK upon tilting from $\theta
=0^{\circ }$ to $10^{\circ }$ as the total magnetic field changes only about
10 percent, which is negligibly small compared to the observed change $10$ K
in $E_{a}$. The in-plane magnetic field can also affect the magnetite of the
exchange energy since the distributions of both the symmetric and
anti-symmetric wavefunctions are expected to vary considerably when the
in-plane magnetic field ranges from 0 at $\theta =0^{\circ }$ to $0.8$ T at $%
\theta =8^{\circ }$. However the sudden change of the activation energy
cannot simply be due to the quantitative change of the exchange energy by
the in-plane magnetic field. First, the experimental data of the energy gap $%
E_{a}$ is nearly constant in a wide range of tilted angle $\theta $ at small
in-plane fields, as shown in Fig.(\ref{energygap}). We therefore expect that
the energy of the pseudospin states is rather insensitive to response
relative to the in-plane magnetic field. Second, we found that such a phase
transition is totally absent at a filling factor $\nu=6$.

We speculate the observations involve a quantum phase transition induced by
an in-plane magnetic field. There is a competition between two ground
states, one of which, at $\theta <\theta _{c}$, takes advantage of Coulomb
interactions by forming pseudospin quantum Hall ferromagnets and the other,
which becomes a more favorable many-body configuration when $\theta >\theta
_{c}$. From the energetic degeneracy of Landau levels before and after the
critical point, we infer that there is a symmetry change: from SU(2) to
SU(4) symmetry in our system. As shown in the standard Landau level fan
diagram (Fig.\ref{phasediagramtilted}f), at $\nu=4$, there is a SU(2)
symmetry at crossing points B and C. Since the two crossing Landau levels
have opposite spins, the exchange energy is highly pseudospin dependent, and
leads to an easy-axis quantum Hall ferromagnet. As a result, the energy gap
here represents an exchange energy penalty for a pseudospin flip. In our
system, as the tilted magnetic field increases, the B and C points are close
to each other, which means a total of four Landau levels with different
subband, orbital, and spin indices are brought close in energy near the
phase transition. Thus there is a SU(4) symmetry at the region where B and C
overlapped. Coupling of all four-fold levels may give rise to a more complex
many-body state. In light of this, the energy gap jump may be due to
suppression of the low energy excitations which originates from the
formation of a new ground state. One possibility is that, in this region,
electrons with different spins and subbands can pair up, condensed in a
pseudospin-singlet pairing state of four-fold Landau levels as $\left\vert
S\right\rangle =\Pi _{k}[C_{(S,1,\uparrow ),k}^{\dagger }+e^{i\varphi
}C_{(A,0,\uparrow ),k}^{\dagger }][C_{(S,1,\downarrow ),k}^{\dagger
}+e^{i\varphi }C_{(A,0,\downarrow ),k}^{\dagger }]\left\vert 0\right\rangle $%
. In this new state, the electrons condense in a superposition of states
with different spins and subbands sharing the same phase $e^{i\varphi }$.
Thus the energy gap of $15$ K for $\theta >\theta _{c}$ can be taken as a
measurement of the pairing strength of pseudospin coherence.

\begin{figure}[h]
\includegraphics[width=0.8\columnwidth]{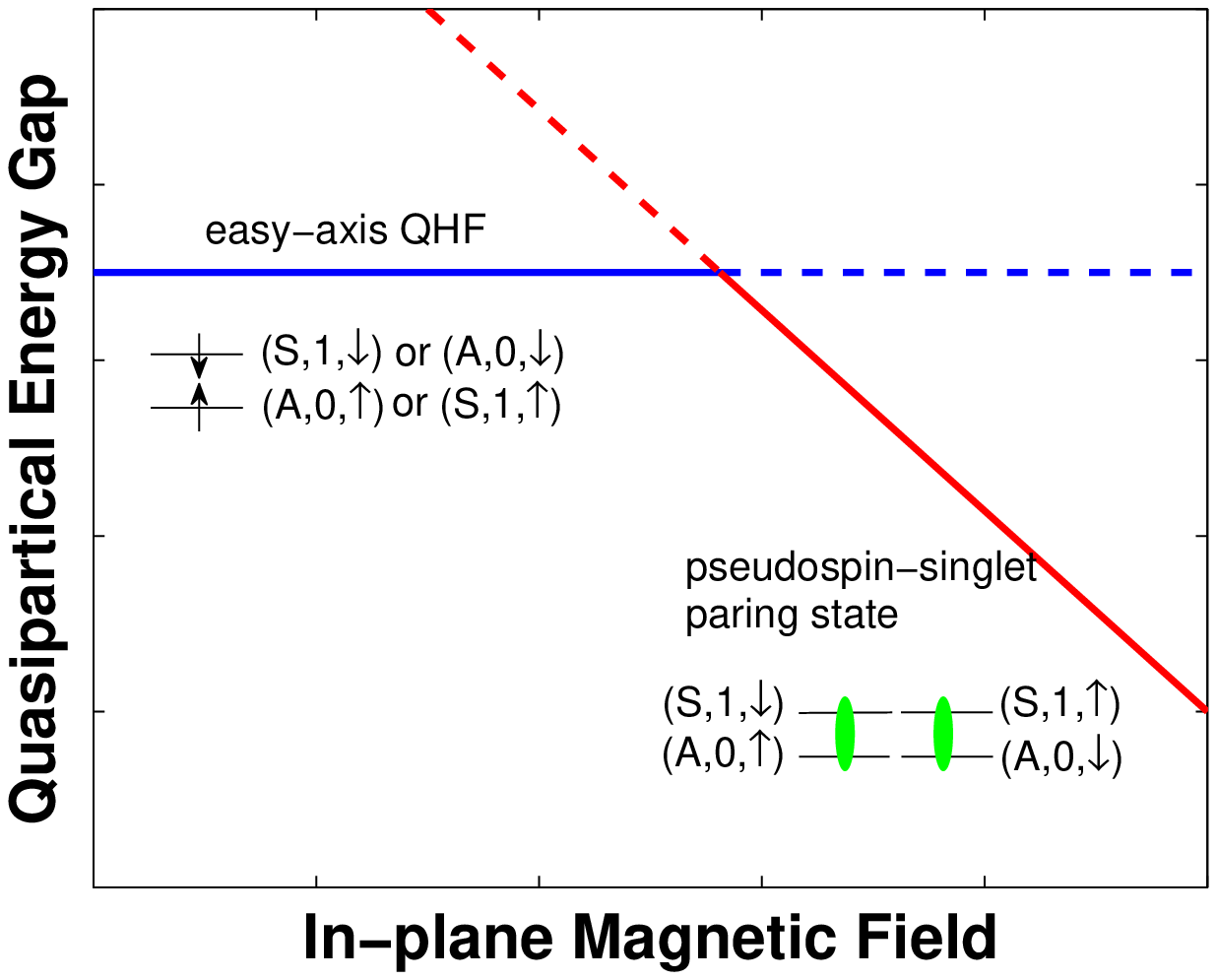}
\caption{Energy profile of the system from two-fold degeneracy to four-fold
degeneracy. The solid line represents the energy of the respective ground
state stabilized. The left is an ordinary easy-axis quantum Hall
ferromagnet(QHF), and the right is a pseudospin-singlet electron pairing
state with different subband and spin indices.}
\label{groundstate}
\end{figure}

As far as we know, there are only a few theoretical investigations of
two-fold and four-fold Landau level degeneracy in two-subband quantum Hall
systems \cite{DasSarma2002,DasSarma2003}. Nevertheless, there are quite a
few theoretical studies of multi-fold degeneracy states in the coupled
bilayer system. A soft barrier originating from Coulomb interaction among
electrons in the quantum well separates our system into effective strongly
coupled bilayers. As a result, the associated wave functions with subband
index can be transformed as those with layer index $\Phi _{L}=[\Phi
_{AS}+\Phi _{S}]/\sqrt{2},\Phi _{R}=[\Phi _{AS}-\Phi _{S}]/\sqrt{2}$. Thus
we believe that the theoretical description of quantum Hall phenomenon in
two-subband single wells only requires slight modifications of the idealized
bilayer model while retaining the basic results of the many-body
Hartree-Fock formalism. On one hand, for an idealized bilayer model at
filling factor $\nu =4$ in zero in-plane magnetic field, where two aligned
Landau levels with different subband and orbital indices are brought into
degeneracy, the ground state can be an easy-axis quantum Hall ferromagnet
dependent on the details of the system \cite{Macdonald2000}. The
quasiparticle energy gap of easy-axis quantum Hall ferromagnet is weakly
dependent on the tilted angle \cite{Macdonald1998}, which is consistent with
our observations for $\theta <\theta _{c}$. On the other hand, for a bilayer
system at $\nu =2$, where sometimes four Landau levels are close to
degeneracy, the ground state is theoretically studied by several groups with
the Hartree-Fock method \cite{Macdonald2002,Demler2004}. It was found that
there are rich phases including ferromagnetic, spin-singlet, and canted
phase and transitions between them. In the presence of strong in-plane
magnetic fields and large tunneling, the ground state is expected to be a
spin-singlet state resulting from the pairing of two adjacent layers\cite%
{Demler2004,DasSarma1997}. In the present system, since the energy
difference between the symmetry and antisymmetry states, $\Delta _{SAS}$, is
as large as $0.75$ $e^{2}/\varepsilon l$, measured from Shubnikov-de Haas
oscillations, one expects a similar pairing state of pseudo-spins, or
sub-bands $\left\vert S\right\rangle $ when four Landau levels are brought
into degeneracy. Thus we can construct a phenomenological model as shown in
the Fig.(\ref{groundstate}), where the ground state is an ordinary easy-axis
quantum Hall ferromagnet for the case of two degenerate Landau levels and
becomes a new state for four degenerate Landau levels, when the electrons
with different spins and subband pair up and coherence associated with the
phase $e^{i\varphi }$ is maintained. We speculate that similar SU(2) to
SU(4) phase transition phenomenon can be observed in bilayer systems at some
appropriate regions. Actually, the activation energies we measured at $%
\theta >\theta _{c}$ are consistent with the experimental values of energy
gap of spin-singlet state in bilayer $\nu =2$, which is about $15$ K\cite%
{Hirayama2006}. Of course, there are possibly different competing-orders,
such as a stripe state with broken translational and spin symmetries, which
is commonly believed to occur at very large in-plane magnetic fields \cite%
{DasSarma2003,Kivelson,Macdonald1999}. More work is desired to calculate the
ground state energy of the proposed SU(4) as a function of the in-plane
field.

In summary, we find experimental evidence for intriguing and unexpected
quantum phase transition from the typical pseudospin quantum Hall
ferromagnetic states with broken SU(2) symmetry into a state with broken
SU(4) symmetry driven by the in-plane magnetic field, around a filling
factor of $\nu=4$ in a two-subband GaAs-AlGaAs 2DEG. While the origin of
this new state is unclear, we discussed the possibility of the pairing state
of electrons with different spins and subbands. More experiments, such as
electrically detected nuclear magnet resonance \cite{Jiang2007}, are needed
to further investigate the microscopic aspects of this quantum phase
transition.

\subsection{Acknowledgement}

The authors would like to thank S. Das Sarma, A. H. Macdonald, K. Yang and
S. A. Kivelson for helpful discussions. This work is supported by National
Fundamental Research Program, the Innovation funds from Chinese Academy of
Sciences, National Natural Science Foundation of China (No.60121503 and
No.10604052).


\begin{thebibliography}{99}
\bibitem{DasSarma} For a review on quantum Hall ferromagnets, see
experimental chapter by Eisenstein, J. P. and theoretical chapter by Girvin,
S. M. and MacDonald, A. H. in \textit{Perspectives on Quantum Hall Effects}
(Wiley, New York, 1997).

\bibitem{Hirayama} Muraki, E., Saku, T. and Hirayama, Y. Charge excitations
in easy-axis and easy-plane quantum Hall ferromagnets. \textit{Phys. Rev.
Lett.} \textbf{87}, 196801 (2001).

\bibitem{Jiang2005} Zhang,X. C., Faulhaber, D. R. and Jiang, H. W. Multiple
phases with the same quantized Hall conductance in a two-subband system.
\textit{Phys. Rev. Lett.} \textbf{95}, 216801 (2005).

\bibitem{Tsui} Lai, K., Pan, W., Tsui, D. C., Lyon, S., Muhlberger, M. and
Schaffler, F. Intervalley gap anomaly of two-dimensional electrons in
silicon. \textit{Phys. Rev. Lett.} \textbf{96}, 076805 (2006).

\bibitem{Shayegan} Vakili, K., Gokmen, T., Gunawan, O., Shkolnikov, Y. P.,
De Pooretere, E. P. and Shayega, M. Dependence of persistent gaps at Landau
level crossings on relative spin. \textit{Phys. Rev. Lett.} \textbf{97}.
116803 (2006).

\bibitem{graphene1} Nomura, K. and Macdonald, A. H. Quantum Hall
ferromagnetism in graphene. \textit{Phys. Rev. Lett.} \textbf{96}, 256602
(2006).

\bibitem{graphene2} Yang, K., Das Sarma, S. and Macdonald, A. H. Collective
modes and skyrmion excitations in graphene SU(4) quantum Hall ferromagnets.
\textit{Phys. Rev. B} \textbf{74}, 075423 (2006).

\bibitem{Wescheider} Piazza, V., Pellegrini, V., Beltram, F., Wescheider,
W., Jungwirth, T. and MacDonald, A. H. First-order phase transitions in a
quantum Hall ferromagnet. \textit{Nature} \textbf{402}, 638 (1999).

\bibitem{Jiang1999} Lee, X. Y., Jiang, H. W. and Schaff, W. J. Topological
phase diagram of a two-subband electron system. \textit{Phys. Rev. Lett.}
\textbf{83}, 3701 (1999).

\bibitem{Jiang2006} Zhang, X. C., Martin, I. and Jiang, H. W. Landau level
anticrossing manifestations in the phase-diagram topology of a two-subband
system. \textit{Phys. Rev. B} \textbf{74}, 073301 (2006).

\bibitem{Macdonald2000} Jungwirth, T. and Macdonald, A. H. Pseudospin
anisotropy classification of quantum Hall ferromagnets. \textit{Phys. Rev. B}
\textbf{63}, 035305 (2000).

\bibitem{Eisenstein} Murphy, S. Q., Eisenstein, J. P., Boebinger, G. S.,
Pfeiffer, L. N. and West, K. W. Many-body integer quantum Hall effect:
Evidence for new phase transitions. \textit{Phys. Rev. Lett.} \textbf{72}.
728 (1994).

\bibitem{DasSarma2002} Wang, D. W., Das Sarma, S., Demler, E. and Halperin,
B. I. Magnetoplasmon excitations and spin density instabilities in an
integer quantum Hall system with a tilted magnetic field. \textit{Phys. Rev.
B} \textbf{66}, 195334 (2002).

\bibitem{DasSarma2003} Wang, D. W., Demler, E. and Das Sarma, S. Spontaneous
symmetry breaking and exotic quantum orders in integer quantum Hall systems
under a tilted magnetic field. \textit{Phys. Rev. B} \textbf{68}, 165303
(2003).

\bibitem{Macdonald1998} Jungwirth, T., Shukla, S. P., Smrcka, L., Shayegan,
M. and Macdonald, A. H. Magnetic Anisotropy in Quantum Hall Ferromagnets.
\textit{Phys. Rev. Lett.} \textbf{81}, 2328 (1998).

\bibitem{Macdonald2002} Burkov, A. A. and Macdonald, A. H. $\nu =2$ bilayer
quantum Hall system in a tilted magnetic field. \textit{Phys. Rev. B}
\textbf{66}, 115323 (2002).

\bibitem{Demler2004} Loptnikova, A., Simon, S. H. and Demler, E. Global
phase diagram of $\nu =2$ quantum Hall bilayers in tilted magnetic fields.
\textit{Phys. Rev. B} \textbf{70}, 115325 (2004).

\bibitem{DasSarma1997} Zheng, L., Radtke, R. J. and Das Sarma, D.
Spin-excitation-instability-induced quantum phase transitions in
double-layer quantum Hall systems. \textit{Phys. Rev. Lett.} \textbf{78},
2453 (1997).

\bibitem{Hirayama2006} Fukuda, A., Sawada, A., Kozumi, S., Terasawa, D.,
Shimoda, Y., Ezawa, Z. F., Kumada, N. and Hirayama, Y. Magnetotransport
study of the canted antiferromagnetic phase in bilayer $\nu =2$ quantum Hall
state. \textit{Phys. Rev. B} \textbf{73}, 165304 (2006).

\bibitem{Kivelson} Fradkin, E. and Kivelson, S. A., Liquid-crystal phases of
quantum Hall systems, \textit{Phys. Rev. B} \textbf{59}, 8065 (1999).

\bibitem{Macdonald1999} Jungwirth, T., MacDonald, A. H., Smrcka, L., and
Girvin, S. M. Field-tilt anisotropy energy in quantum Hall stripe state,
\textit{Phys. Rev. B} \textbf{60}, 15574 (1999)

\bibitem{Jiang2007} Zhang, X. C., Scott, G. D. and Jiang, H. W. NMR probing
of spin excitations in the ring structure of a two-subband system. \textit{%
Phys. Rev. Lett.} \textbf{98}, 246802 (2007).
\end{thebibliography}
\end{document}